\documentclass[journal,12pt,onecolumn,draftclsnofoot,]{IEEEtran}
\usepackage{amsmath,amsfonts}
\usepackage{color}
\usepackage{algorithmic}
\usepackage{algorithm}
\usepackage{array}
\usepackage[caption=false,font=normalsize,labelfont=sf,textfont=sf]{subfig}
\usepackage{textcomp}
\usepackage{stfloats}
\usepackage{url}
\usepackage{verbatim}
\usepackage{graphicx}
\usepackage{cite}

\newtheorem{Theorem}{\textbf{Theorem}}
\newtheorem{Lemma}{\textbf{Lemma}}

\begin{document}
\title{Secret-Key Capacity from MIMO Channel Probing}
\author{Yingbo Hua,~\IEEEmembership{Fellow,~IEEE}, and Ahmed Maksud,~\IEEEmembership{Member,~IEEE}
\thanks{This paper is in press for publication in IEEE Wireless Communications Letters. \copyright 2024 IEEE. Personal use of this material is permitted. Permission from IEEE must be
obtained for all other uses, in any current or future media, including
reprinting/republishing this material for advertising or promotional purposes, creating new
collective works, for resale or redistribution to servers or lists, or reuse of any copyrighted
component of this work in other works. The authors are with Department of Electrical and Computer Engineering, University of California at Riverside,
Riverside, CA, USA. Emails: yhua@ece.ucr.edu and
ahmed.maksud@email.ucr.edu. This work was supported in part by the U.S. Department of Defense under the Agreement W911NF-20-2-0267. The views and conclusions contained in this
document are those of the authors and should not be interpreted as representing the official policies, either
expressed or implied, of the U.S. Government. The U.S. Government is
authorized to reproduce and distribute reprints for Government purposes notwithstanding any copyright
notation herein.}}

%
%
%
%
%
%
\maketitle
\begin{abstract}
Revealing expressions of secret-key capacity (SKC) based on data sets from Gaussian MIMO channel probing are presented. It is shown that Maurer's upper and lower bounds on SKC coincide when the used data sets are produced from one-way channel probing. As channel coherence time increases, SKC in bits per probing channel use is always lower bounded by a positive value unless eavesdropper's observations are noiseless, which is unlike SKC solely based on reciprocal channels.
\end{abstract}
\begin{IEEEkeywords}
Physical layer security, secret-key generation, secret-message transmission.
\end{IEEEkeywords}
\section{Introduction}\label{introduction}

A central problem of physical layer security (PLS) is for two friendly nodes (Alice and Bob) to exchange a secret message against an eavesdropper (Eve). There are two primary approaches to the PLS problem: direct transmission of a secret message from Alice to Bob (or in reverse direction); and establishment of a secret key between Alice and Bob (so that it can be used to protect future transmissions). The former is also known as wiretap channel (WTC) problem while the latter as secret key generation (SKG) problem. Good reviews on PLS are available in \cite{Zhang2020}, \cite{Poor2017}, \cite{Bloch2011} among others.

Given a system of Gaussian MIMO channels between Alice, Bob and Eve, the WTC approach has been widely studied and facilitated by revealing expressions of its secrecy capacity (directly in terms of the channel matrices) established in \cite{Khisti2010} and \cite{Oggier2011}. Given the same system of channels, the SKG approach is also applicable but had received less thorough investigations.


The first and most crucial step for SKG out of such a system is to generate correlated data sets at Alice and Bob (before information reconciliation and privacy amplification are conducted for secret key agreement \cite{Zhang2020}). Assuming that the generated (random) data sets $\mathcal{X}$, $\mathcal{Y}$ and $\mathcal{Z}$ at Alice, Bob and Eve are memoryless, the secret-key capacity (SKC) based on $\{\mathcal{X};\mathcal{Y};\mathcal{Z}\}$ is subject to known expressions of its lower and upper bounds as established in \cite{Maurer1993} and \cite{Ahlswede1993}.

Based on $\{\mathcal{X};\mathcal{Y};\mathcal{Z}\}$ generated by random channel probing over the MIMO channels, the recent work \cite{Hua2023} established simple expressions of the degree-of-freedom (DoF) of SKC. Further study is shown in \cite{MaksudHua2023} and \cite{Hua2023SISO}. But no exact expression of the SKC for any MIMO channels was available until this work. The main contribution of this paper is shown in Theorem \ref{Theorem_a} in section \ref{SKC}. The fundamentals of information theory from \cite{Cover2006} are used extensively.

\section{MIMO Channel Probing and Data Model}\label{System_Model}
We consider a MIMO channel between two legitimate nodes A and B (Alice and Bob) in the presence of an Eavesdropper (Eve). The numbers of antennas on these nodes are respectively $n_A$, $n_B$ and $n_E$. The channel response matrices from Alice to Bob and from Bob to Alice are denoted by $\mathbf{H}_{BA}$ and $\mathbf{H}_{AB}$ respectively, and the channel response matrices from Alice to Eve and from Bob to Eve are denoted by $\mathbf{G}_A$ and $\mathbf{G}_B$ respectively. Note that all channels are flat-fading within the bandwidth or subcarrier of interest. Also note that all channels are assumed to be block-wise fading, i.e., all channel matrices are constant within each coherence period but vary independently from one coherence period to another.

The channel probing scheme considered in this paper is as follows.
Each of the channel coherence periods is divided into four windows. In window 1, Alice transmits a row-wise orthogonal public pilot matrix $\sqrt{\alpha_A P}\mathbf{\Pi}_A\in\mathcal{C}^{n_A\times \phi_A}$ over $n_A$ antennas and $\phi_A$ time slots where $\mathbf{\Pi}_A\mathbf{\Pi}_A^H=\psi_A\mathbf{I}_{n_A}$. In other words, the $i$th row of $\sqrt{\alpha_A P}\mathbf{\Pi}_A$ is transmitted from the $i$th antenna of Alice, and the $j$th column of $\sqrt{\alpha_A P}\mathbf{\Pi}_A$ is transmitted by Alice in time slot $j$ of window 1.   In window 2, Alice transmits a random matrix $\sqrt{\alpha_A P}\mathbf{X}_A\in\mathcal{C}^{n_A\times v_A}$  over $n_A$ antennas and $v_A$ time slots. Similarly, in window 3, Bob transmits  a row-wise orthogonal public pilot matrix $\sqrt{\alpha_B P}\mathbf{\Pi}_B\in\mathcal{C}^{n_B\times \phi_B}$ where $\mathbf{\Pi}_B\mathbf{\Pi}_B^H=\psi_B\mathbf{I}_{n_B}$. And in window 4, Bob transmits a random matrix $\sqrt{\alpha_B P}\mathbf{X}_B\in\mathcal{C}^{n_B\times v_B}$ over $n_B$ antennas and $v_B$ time slots.

The above probing scheme is a two-way half-duplex scheme and a special case among those considered in \cite{Hua2023} where DoF of SKC is presented. This scheme differs from the earlier schemes in \cite{Aldaghri2020} and \cite{Li2022a} where no public pilot is used while reciprocal channel is required.

All entries in the random matrices $\mathbf{X}_A$ and $\mathbf{X}_B$ will have the unit variance. If each entry in $\mathbf{\Pi}_A$ and $\mathbf{\Pi}_B$ also has the unit power, then $\psi_A=\phi_A$ and $\psi_B=\phi_B$. In general, we have $\phi_A\geq n_A$ and $\phi_B\geq n_B$, which is necessary for both $\mathbf{\Pi}_A$ and $\mathbf{\Pi}_B$ to be each row-wise orthogonal.

The (nominal) transmit power by Alice from each antenna in each slot is represented by $\alpha_A P$, and that by Bob is represented by $\alpha_B P$.
We will assume $\psi_A\gg n_A$ and $\psi_B\gg n_B$ so that the channel estimation errors at all nodes based on the public pilots are negligible as explained later.

The signals received by Bob in windows 1 and 2 are represented by $\mathbf{Y}_{B}^{(1)}\in \mathcal{C}^{n_B\times \phi_A }$ and $\mathbf{Y}_{B}^{(2)}\in \mathcal{C}^{n_B\times v_A}$ respectively. We also write $\mathbf{Y}_{B}\doteq[\mathbf{Y}_{B}^{(1)},\mathbf{Y}_{B}^{(2)}]$.

The signals received by Alice in windows 3 and 4 are denoted by $\mathbf{Y}_{A}^{(1)}\in \mathcal{C}^{n_A\times \phi_B}$ and $\mathbf{Y}_{A}^{(2)}\in \mathcal{C}^{n_A\times v_B}$. Also let $\mathbf{Y}_{A}\doteq[\mathbf{Y}_{A}^{(1)},\mathbf{Y}_{A}^{(2)}]$.

The signals received by Eve in windows 1, 2, 3 and 4 are  respectively  $\mathbf{Y}_{EA}^{(1)}\in \mathcal{C}^{n_E\times \phi_A }$, $\mathbf{Y}_{EA}^{(2)}\in \mathcal{C}^{n_E\times  v_A}$, $\mathbf{Y}_{EB}^{(1)}\in \mathcal{C}^{n_E\times \phi_B}$ and $\mathbf{Y}_{EB}^{(2)}\in \mathcal{C}^{n_E\times v_B}$. Also let
$\mathbf{Y}_{EA}\doteq[\mathbf{Y}_{EA}^{(1)},\mathbf{Y}_{EA}^{(2)}]$ and
$\mathbf{Y}_{EB}\doteq[\mathbf{Y}_{EB}^{(1)},\mathbf{Y}_{EB}^{(2)}]$.

Note that the matrices with the superscript $^{(1)}$ are associated with the public pilots, and those with $^{(2)}$ are associated with the random symbols. More specifically, we can write
\begin{subequations}
\begin{align}
&\mathbf{Y}_{A}=\sqrt{\gamma_{AB}}\left[\mathbf{H}_{AB}\mathbf{\Pi}_B,\mathbf{H}_{AB}\mathbf{X}_B\right]+
    \mathbf{W}_{A},\label{y_A}\\
    &\mathbf{Y}_{B}=\sqrt{\gamma_{BA}}\left[\mathbf{H}_{BA}\mathbf{\Pi}_A,\mathbf{H}_{BA}\mathbf{X}_A\right]
    +\mathbf{W}_{B},\label{y_B}\\
     &\mathbf{Y}_{EA}=\sqrt{\gamma_{EA}}\left[\mathbf{G}_{A}\mathbf{\Pi}_A,\mathbf{G}_{A}\mathbf{X}_A\right]
     +
    \mathbf{W}_{EA},\label{y_EA}\\
    &\mathbf{Y}_{EB}=\sqrt{\gamma_{EB}}\left[\mathbf{G}_{B}\mathbf{\Pi}_B,\mathbf{G}_{B}\mathbf{X}_B\right]
     +
    \mathbf{W}_{EB}\label{y_EB}.
\end{align}
\end{subequations}
Here all entries in the normalized noise matrices (i.e., the $\mathbf{W}$ matrices) have the unit variance. We have used $\gamma_{AB}=\frac{\alpha_B P}{\lambda_A}$  where $\lambda_A$ is the noise variance at Alice after the normalized $\mathbf{H}_{BA}$ (as shown later) but before the normalized $\mathbf{W}_A$. This definition of noise variance at Alice is applied similarly for other nodes.  Namely
$\gamma_{BA}=\frac{\alpha_A P}{\lambda_B}$
where $\lambda_B$ is the noise variance at Bob. Furthermore, $\gamma_{EA}=\frac{\alpha_A P}{\lambda_{EA}}$ and $\gamma_{EB}=\frac{\alpha_B P}{\lambda_{EB}}$ where $\lambda_{EA}$ is noise variance at Eve relative to the channel from Alice  and $\lambda_{EB}$ is noise variance at Eve relative to the channel from Bob. Since the receive channel gains at Eve relative to Alice and Bob are different from each other in general, we have $\lambda_{EA}\neq\lambda_{EB}$ in general even if the actual noise (such as thermal noise) at Eve has the same variance at all times. For example, if Eve is closer in distance to Alice than to Bob, then we should expect $\lambda_{EA}<\lambda_{EB}$.

All entries in $\mathbf{X}_A\in\mathcal{C}^{n_A\times v_A}$, $\mathbf{X}_B\in\mathcal{C}^{n_B\times v_B}$, $\mathbf{H}_{BA}\in\mathcal{C}^{n_B\times n_A}$ (or $\mathbf{H}_{AB}\in\mathcal{C}^{n_A\times n_B}$), $\mathbf{G}_{A}\in\mathcal{C}^{n_E\times n_A}$,  $\mathbf{G}_{B}\in\mathcal{C}^{n_E\times n_B}$ and all the $\mathbf{W}$ matrices are normalized to be i.i.d. $\mathcal{CN}(0,1)$. The simulation results shown later are based on $10^4$ independent realizations of these entries.

We will treat $\mathbf{H}_{BA}$ and $\mathbf{H}_{AB}$ as jointly Gaussian with the correlation matrix  $\mathbb{E}\{\mathbf{h}_{AB}^t\mathbf{h}_{BA}^H\}=\rho\mathbf{I}_{n_An_B}$. Here $\mathbf{h}_{BA}=vec(\mathbf{H}_{BA})$ and $\mathbf{h}_{AB}^t=vec(\mathbf{H}_{AB}^T)$. Let $\mathbf{C}_{\mathbf{x}|\mathbf{y}}$ denote the conditional covariance matrix of $\mathbf{x}$ given $\mathbf{y}$. It follows that $\mathbf{C}_{\mathbf{h}_{AB}|\mathbf{h}_{BA}^t}=\mathbf{C}_{\mathbf{h}_{BA}^t|\mathbf{h}_{AB}}
=(1-|\rho|^2)\mathbf{I}_{n_An_B}$. Here, $|\rho|=1$ if all channel parameters between Alice and Bob are perfectly reciprocal, and $|\rho|<1$ if every channel parameter between Alice and Bob is not perfectly reciprocal.

After the previously described channel probing, the (random) data sets $\mathcal{X}$, $\mathcal{Y}$ and $\mathcal{Z}$ available at Alice, Bob and Eve respectively  in each coherence period are as follows: $\mathcal{X}=
    \left\{\mathbf{X}_A,\mathbf{Y}_{A}^{(1)},\mathbf{Y}_{A}^{(2)}\right\}$; $\mathcal{Y}=
    \left\{\mathbf{X}_B,\mathbf{Y}_{B}^{(1)},\mathbf{Y}_{B}^{(2)}\right\}$; $\mathcal{Z}=
    \left\{\mathbf{Y}_{EA}^{(1)},\mathbf{Y}_{EA}^{(2)},\mathbf{Y}_{EB}^{(1)},
    \mathbf{Y}_{EB}^{(2)}\right\}$.
%
%

 Let $C_A\doteq I(\mathcal{X};\mathcal{Y})-I(\mathcal{X};\mathcal{Z})=h(\mathcal{X}|\mathcal{Z})-
 h(\mathcal{X}|\mathcal{Y})$, $C_B\doteq I(\mathcal{X};\mathcal{Y})-I(\mathcal{Y};\mathcal{Z})=h(\mathcal{Y}|\mathcal{Z})-
 h(\mathcal{Y}|\mathcal{X})$, and $C_Z\doteq I(\mathcal{X};\mathcal{Y}|\mathcal{Z})
 =h(\mathcal{X}|\mathcal{Z})-h(\mathcal{X}|\mathcal{Y},\mathcal{Z})$. It follows from \cite{Maurer1993} and \cite{Ahlswede1993} (and also the generalized mutual information \cite{Cover2006}) that the secret-key capacity $C_S$ (in bits per coherence period) based on $\mathcal{X}$, $\mathcal{Y}$ and $\mathcal{Z}$ satisfies $\max(C_A,C_B)\leq C_S\leq C_Z$.

 It follows from \cite{Hua2023} that for $n_A\geq n_B$ and  relative to $\log P$, $\texttt{DoF}(C_A)\leq \texttt{DoF}(C_B)=\texttt{DoF}(C_S)=\texttt{DoF}(C_Z)$. This suggests that if  $n_A\geq n_B$, the gap between $C_B$ and $C_Z$ should be small at high power. Note: $\texttt{DoF}(C)\doteq\lim_{P\to\infty}\frac{C}{\log P}$.

%
\section{Secret-Key Capacity from MIMO Probing}\label{SKC}

The following lemmas will be needed.
\begin{Lemma}\label{Lemma_high_power}
Let $\mathbf{Y}=\sqrt{\gamma}\mathbf{H}\mathbf{\Pi}+\mathbf{W}$ and $\mathbf{Y}'=\sqrt{\gamma}\mathbf{H}\mathbf{X}+\mathbf{W}'$ with $\mathbf{H}\in\mathcal{C}^{N\times K}$, $\mathbf{\Pi}\in\mathcal{C}^{K\times \phi}$, $\phi\geq K$, $\mathbf{\Pi}\mathbf{\Pi}^H=\psi\mathbf{I}_{K}$, and all entries in $\mathbf{H}$, $\mathbf{W}$, $\mathbf{X}$ and $\mathbf{W}'$ being i.i.d. $\mathcal{CN}(0,1)$. Then for $\gamma\geq 1$ and  $\psi\gg K$, the effect of the errors in the optimal estimate of $\mathbf{H}$ from $\mathbf{Y}$ and $\mathbf{\Pi}$ on  $\mathbf{Y}'$ is negligible. In other words, given $\mathbf{Y}$, $\mathbf{\Pi}$ and a large $\psi$, we can treat $\mathbf{H}$ as known in dealing with $\mathbf{Y}'$.
\end{Lemma}
\begin{IEEEproof}
This is easy to prove.
\end{IEEEproof}

\begin{Lemma}\label{Lemma_2}
  Let $\mathbf{Y}=\sqrt{\gamma}\mathbf{HX}+\mathbf{W}$ with $\mathbf{H}\in\mathcal{C}^{N\times K}$, $\mathbf{X}\in\mathcal{C}^{K\times M}$ and all entries in $\mathbf{X}$ and $\mathbf{W}$ being i.i.d. $\mathcal{CN}(0,1)$. Then $h(\mathbf{Y}|\mathbf{H})=NM\log(\pi e)+M\mathbb{E}\{\log|\gamma\mathbf{H}\mathbf{H}^H+\mathbf{I}_{N}|\}$ and
$I(\mathbf{Y};\mathbf{X}|\mathbf{H})=
M\mathbb{E}\{\log|\gamma\mathbf{H}\mathbf{H}^H+\mathbf{I}_{N}|\}
$.
\end{Lemma}
\begin{IEEEproof}
This is a known result, e.g., see \cite{Hua2023}.
\end{IEEEproof}

\begin{Lemma}\label{Lemma_MI}
Recall $\mathbf{Y}_A^{(1)}$ and $\mathbf{Y}_B^{(1)}$ defined in section \ref{System_Model}. Then
\begin{equation}\label{eq:CS1}
  C_S^{(1)}\doteq I(\mathbf{Y}_A^{(1)};\mathbf{Y}_B^{(1)})=n_An_B\log g
\end{equation}
with
\begin{equation}\label{eq:g}
  g=\frac{(\gamma_{AB}\psi_B+1)
(\gamma_{BA}\psi_A+1)}{(1-|\rho|^2)\gamma_{AB}\psi_B\gamma_{BA}\psi_A+\gamma_{AB}\psi_B
+\gamma_{BA}\psi_A+1}.
\end{equation}
\end{Lemma}
\begin{IEEEproof}
See Appendix\ref{Lemma_MI_proof}.
\end{IEEEproof}

It is important to add a remark here in dealing with (for example) $I(\mathbf{Y}_A^{(1)};\mathbf{Y}_B^{(1)})=h(\mathbf{Y}_A^{(1)})-
h(\mathbf{Y}_A^{(1)}|\mathbf{Y}_B^{(1)})$. Lemma \ref{Lemma_high_power} implies that for a large $\psi_A$, a given $\mathbf{Y}_B^{(1)}$ implies a given $\mathbf{H}_{BA}$, i.e., $h(\mathbf{Y}_A^{(1)}|\mathbf{Y}_B^{(1)})\approx h(\mathbf{Y}_A^{(1)}|\mathbf{Y}_B^{(1)},\mathbf{H}_{BA})$. But here $h(\mathbf{Y}_A^{(1)}|\mathbf{Y}_B^{(1)})\not\approx h(\mathbf{Y}_A^{(1)}|\mathbf{H}_{BA})$ due to correlation between $\mathbf{Y}_A^{(1)}$ and $\mathbf{Y}_B^{(1)}$ even when $\mathbf{H}_{BA}$ is given. However, we will use frequently such approximation $h(\mathbf{Y}'|\mathbf{Y}_B^{(1)})\approx h(\mathbf{Y}'|\mathbf{H}_{BA})$ for a large $\psi_A$ where $\mathbf{Y}'$ and $\mathbf{Y}_B^{(1)}$ are independent of each other when conditioned on $\mathbf{H}_{BA}$. For all approximations that hold under given conditions, we will also use ``$\approx$'' and ``$=$'' interchangeably

\begin{Theorem}\label{Theorem_a}
  Assume large $\psi_A$ and $\psi_B$, and any $n_A\geq 1$, $n_B\geq 1$ and $n_E\geq 1$. The gap between $C_Z$ and $C_B$ is
  \begin{align}
&C_Z-C_B =v_B\mathbb{E}\{\log|\mathbf{I}_{n_B}+
    \gamma_{AB}\tilde{\mathbf{H}}_{AB}^H\tilde{\mathbf{H}}_{AB}|\}\notag\\
    &\,\,-v_B\mathbb{E}\{\log|\mathbf{I}_{n_B}+\gamma_{AB}\mathbf{H}_{AB}^H\mathbf{H}_{AB}|\}
\end{align}
where
$
  \tilde{\mathbf{H}}_{AB}^H\tilde{\mathbf{H}}_{AB}
  =\mathbf{H}_{AB}^H\mathbf{H}_{AB}+\frac{\lambda_A}{\lambda_{EB}}\mathbf{G}_B^H\mathbf{G}_B
$.
Equivalently,
 \begin{align}\label{eq:diff1}
 &C_Z-C_B=v_B\mathbb{E}\left \{\log\left |\mathbf{I}_{n_B}+\gamma_{AB}
 \frac{\lambda_A}{\lambda_{EB}}\mathbf{G}_B^H\mathbf{G}_B\right .\right .\notag\\
 &\,\,\cdot\left .\left .\left (\mathbf{I}_{n_B}+\gamma_{AB}\mathbf{H}_{AB}^H\mathbf{H}_{AB}\right )^{-1}
 \right |\right \}\geq 0
 \end{align}
 with equality if and only if $v_B=0$ (provided $\gamma_{AB}>0$ and $\frac{\lambda_A}{\lambda_{EB}}>0$).
Furthermore,
\begin{align}\label{eq:C_B_general_1}
&C_B=C_S^{(1)}+v_A\xi_B-v_B\mathbb{E}\{\log|\gamma_{EB}\mathbf{G}_B^H\mathbf{G}_B+\mathbf{I}_{n_B}|\}\notag\\
&\,\,+v_B\mathbb{E}\{\log|\gamma_{AB}\mathbf{H}_{AB}^H\mathbf{H}_{AB}+\mathbf{I}_{n_B}|\}
\end{align}
with
\begin{align}
&\xi_B=\mathbb{E}\{\log|\gamma_{EA}\tilde{\mathbf{G}}_{A}^H\tilde{\mathbf{G}}_{A}+\mathbf{I}_{n_A}|\}
\notag\\
&\,\,-\mathbb{E}\{\log|\gamma_{EA}\mathbf{G}_A^H\mathbf{G}_A+\mathbf{I}_{n_A}|\}
\end{align}
and
$
  \tilde{\mathbf{G}}_{A}^H\tilde{\mathbf{G}}_{A}=
  \mathbf{G}_{A}^H\mathbf{G}_{A}+\frac{\lambda_{EA}}{\lambda_B}
    \mathbf{H}_{BA}^H\mathbf{H}_{BA}
$. Equivalently,
\begin{align}
&\xi_B=\mathbb{E}\left \{\log\left |\mathbf{I}_{n_A}+\gamma_{BA}
    \mathbf{H}_{BA}^H\mathbf{H}_{BA}\right . \right .\notag\\
&\,\,\left .\left . \cdot
    \left (\gamma_{BA}(\lambda_B/\lambda_{EA})\mathbf{G}_A^H\mathbf{G}_A+\mathbf{I}_{n_A}\right )^{-1}\right |\right\} \geq 0
\end{align}
with equality only if $\frac{\lambda_B}{\lambda_{EA}}=\infty$ (provided $\gamma_{BA}>0$).
\end{Theorem}
\begin{IEEEproof}
See Appendix\ref{proof_of_theorem_a}.
\end{IEEEproof}

\subsection{Discussion of Theorem \ref{Theorem_a}}

Theorem \ref{Theorem_a} does not require $n_A\geq n_B$. But if $n_A\geq n_B$, we see that both $\mathbf{H}_{AB}$ and $\tilde{\mathbf{H}}_{AB}$ have the full column rank $n_B$ for all $n_E\geq 1$ and hence (one can verify) $\texttt{DoF}(C_Z-C_B)=0$ for all $v_A\geq 1$, $v_B\geq 1$ and $n_E\geq 1$. This is consistent with a previous result shown in \cite{Hua2023}.

If $v_A\geq 1$ and $v_B=0$ (i.e., one-way channel probing from Alice to Bob), then $C_B=C_Z$ and hence
\begin{align}\label{}
  &\frac{1}{v_A}C_S=\frac{1}{v_A}C_B = \frac{1}{v_A}C_Z=\frac{1}{v_A}C_S^{(1)}+\xi_B\geq \xi_B
\end{align}
with equality if $\rho=0$ or $v_A\to\infty$.
Since Theorem \ref{Theorem_a} does not require $n_A\geq n_B$, it also follows that if $v_A=0$ and $v_B\geq 1$ then $C_S=C_A=C_Z$ (by symmetry between $C_A$ and $C_B$). In other words, if the channel probing is done only in one direction, the secret-key capacity $C_S$ based on the corresponding data sets always coincides with the corresponding Maurer's lower and upper bounds.

But the channel probing from a node with more antennas to another node with less antennas should generally result in a larger $C_S$ in the regime of high power. This is because
for $n_A\geq n_B$,
$\texttt{DoF}(C_S)=v_A\min[n_B,(n_A-n_E)^+]
    +v_B\left(n_B-n_E\right)^+ +\delta_\rho n_An_B
$
\cite{Hua2023} where $\delta_\rho=1$ if $|\rho|=1$, and $\delta_\rho=0$ if $|\rho|<1$.
Then subject to $v_A+v_B\leq v^*$, $\texttt{DoF}(C_S)$ is maximized by $v_A=v^*$ and $v_B=0$.

Theorem \ref{Theorem_a} also implies that for one-way channel probing from Alice to Bob, the resulting secret-key capacity $\frac{C_S}{v_A}$ in bits per probing instant is always lower bounded by $\xi_B$ which is positive as long as $\lambda_{EA}>0$ (i.e., the signals received by Eve from Alice are not noiseless).

\begin{figure}[ht]
\begin{minipage}[b]{0.48\linewidth}
\centering
\includegraphics[width=\textwidth]{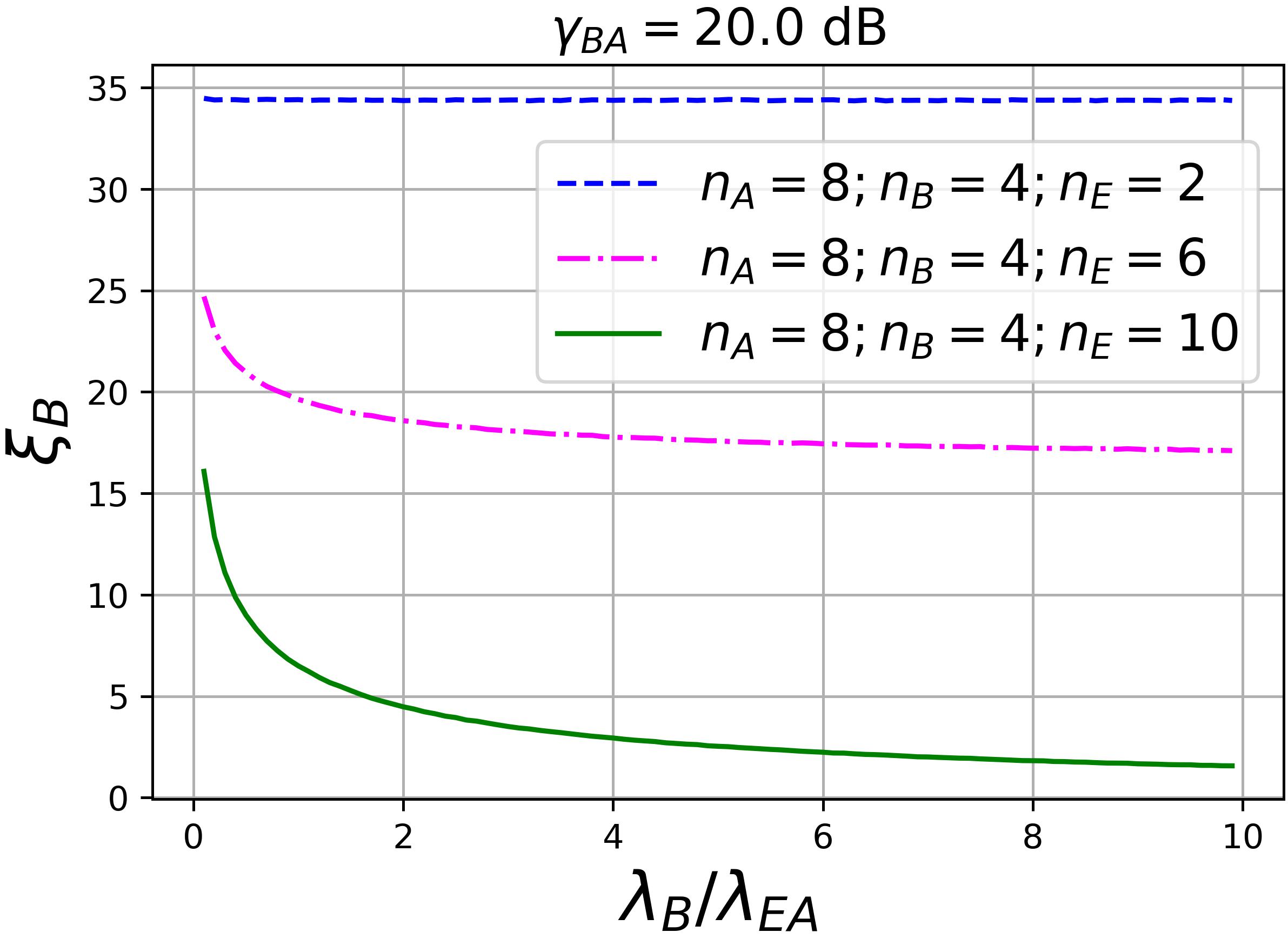}
\caption{$\xi_B$ vs $\lambda_B/\lambda_{EA}$.}
\label{fig:fig_xiB_A}
\end{minipage}
\hspace{0.1cm}
\begin{minipage}[b]{0.48\linewidth}
\centering
\includegraphics[width=\textwidth]{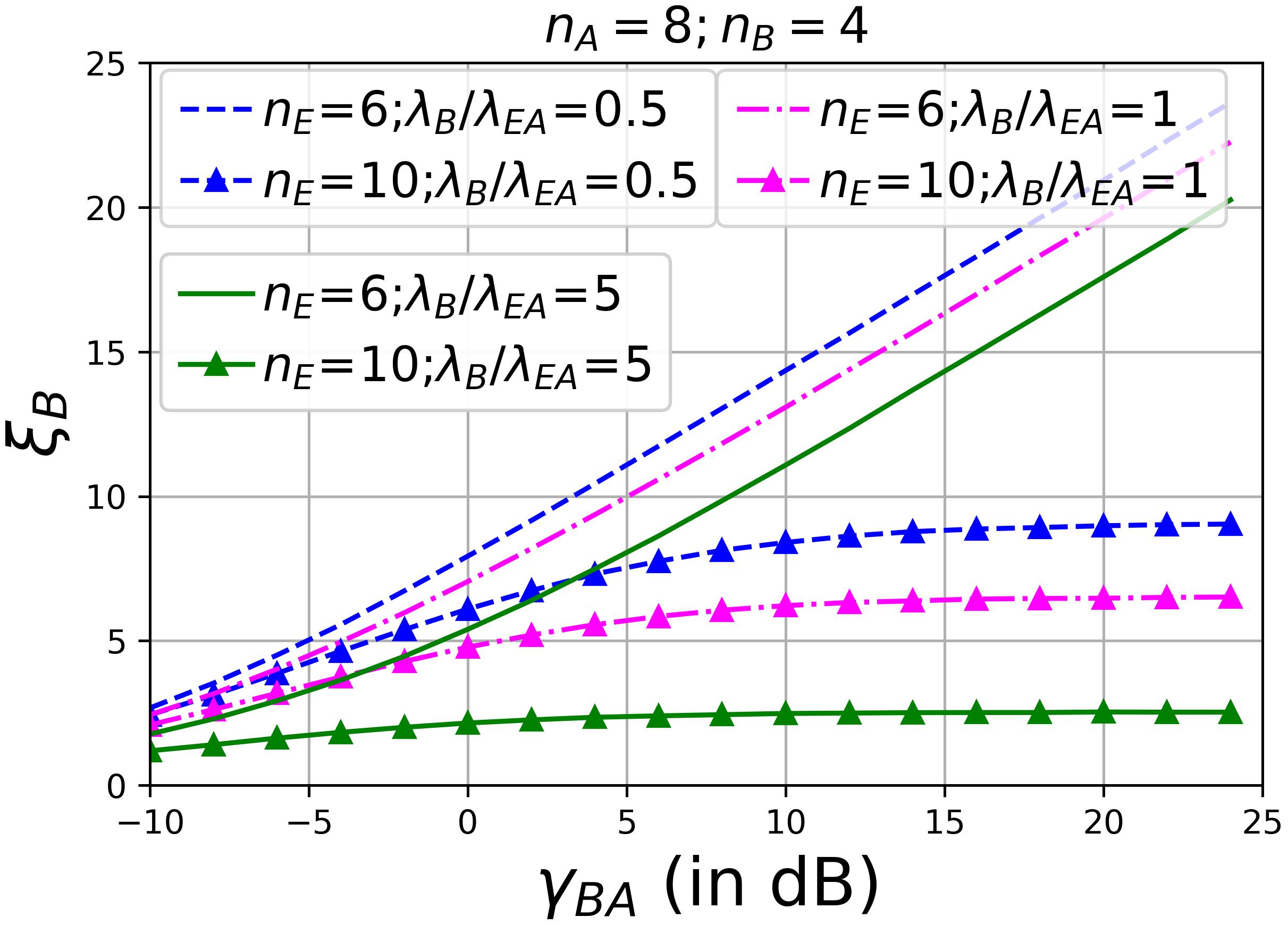}
\caption{$\xi_B$ vs $\gamma_{BA}=\alpha_A P/\lambda_B$.}
\label{fig:fig_xiB_B}
\end{minipage}
\end{figure}

Numerical illustrations of  $\xi_B$ are shown in Figs \ref{fig:fig_xiB_A} and \ref{fig:fig_xiB_B}. Fig. \ref{fig:fig_xiB_A} illustrates $\xi_B>0$ in all cases under $\lambda_B/\lambda_{EA}<\infty$. Fig. \ref{fig:fig_xiB_B} confirms the theory $\texttt{DoF}(\xi_B)=\min[n_B,(n_A-n_E)^+]$; i.e., $\texttt{DoF}(\xi_B)\doteq\lim_{P\to\infty}\frac{\xi_B}{\log P}=2$ for $n_A=8$, $n_B=4$ and $n_E=6$, and $\texttt{DoF}(\xi_B)=0$ for $n_A=8$, $n_B=4$ and $n_E=10$.


The contribution of $v_B>0$ to $C_B$ is either positive or negative, depending on whether or not $|\gamma_{AB}\mathbf{H}_{AB}^H\mathbf{H}_{AB}+\mathbf{I}_{n_B}|>
|\gamma_{EB}\mathbf{G}_B^H\mathbf{G}_B+\mathbf{I}_{n_B}|$, i.e.,
whether or not the MIMO capacity from Bob to Alice is larger than that from Bob to Eve (subject to uniform power scheduling).

\section{Conclusion}
For the first time, closed-form expressions of SKC based on data sets from a Gaussian MIMO channel probing are shown. The gap between Maurer's upper and lower bounds is proven to be zero when the data sets used are from one-way probing. Furthermore, it is now established that SKC in bits per second from channel probing is not constrained by channel coherence time, which is unlike SKC based on reciprocal channel responses. These results are complementary to the prior works on DoF of SKC from MIMO channel probing. Compared to quantum key distribution \cite{QKD2018}, SKG from radio or any non-quantum channels is much more cost-effective. Theorem 1 provides a strong motivation for further development of radio or non-quantum based schemes for SKG.

\section*{Appendix}
\begin{appendices}

\subsection{Proof of Lemma \ref{Lemma_MI}}\label{Lemma_MI_proof}
We can write:
\begin{align}\label{eq:Lemma_MI_proof_1}
    I(\mathbf{Y}_A^{(1)};\mathbf{Y}_B^{(1)})=h(\mathbf{Y}_A^{(1)})+
    h(\mathbf{Y}_B^{(1)})-h(\mathbf{Y}_A^{(1)},\mathbf{Y}_B^{(1)}).
\end{align}
It follows from \eqref{y_A} and \eqref{y_B} that
\begin{subequations}\label{eq:Lemma_MI_proof_2}
\begin{align}
    \mathbf{y}_A^{(1)}&\doteq vec(\mathbf{Y}_A^{(1)})=\sqrt{\gamma_{AB}}(\mathbf{\Pi}_B^T\otimes\mathbf{I}_{n_A})
    \mathbf{h}_{AB}+\mathbf{w}_A^{(1)},\notag\\
    \mathbf{y}_B^{(1)t}&\doteq vec(\mathbf{Y}_B^{(1)T})=\sqrt{\gamma_{BA}}(\mathbf{I}_{n_B}\otimes
    \mathbf{\Pi}_A^T)\mathbf{h}_{BA}^t+
    \mathbf{w}_B^{(1)t},\notag
\end{align}
\end{subequations}
\begin{equation}\label{}
  \mathbf{A}\doteq\mathbb{E}\{\mathbf{y}_A^{(1)}\mathbf{y}_A^{(1)H}\}=
  \gamma_{AB}(\mathbf{\Pi}_B^T\mathbf{\Pi}_B^*\otimes\mathbf{I}_{n_A})+\mathbf{I}_{n_A\phi_B},
\end{equation}
\begin{equation}\label{}
  \mathbf{B}\doteq\mathbb{E}\{\mathbf{y}_B^{(1)}\mathbf{y}_B^{(1)tH}\}=
  \gamma_{BA}(\mathbf{I}_{n_B}\otimes\mathbf{\Pi}_A^T\mathbf{\Pi}_A^*)+\mathbf{I}_{n_B\phi_A},
\end{equation}
\begin{equation}\label{}
  \mathbf{C}\doteq\mathbb{E}\{\mathbf{y}_A^{(1)}\mathbf{y}_B^{(1)tH}\}=\rho
  \sqrt{\gamma_{BA}\gamma_{AB}}(\mathbf{\Pi}_B^T\otimes\mathbf{\Pi}_A^*).
\end{equation}
Then it follows from \eqref{eq:Lemma_MI_proof_1} that
\begin{align}
&I(\mathbf{Y}_A^{(1)};\mathbf{Y}_B^{(1)})
=\log|\mathbf{A}|+\log|\mathbf{B}|-\log\left |\begin{array}{cc}
                                                \mathbf{A} & \mathbf{C} \\
                                                \mathbf{C}^H & \mathbf{B}
                                              \end{array}
\right |.
\end{align}

We will use $\left |\begin{array}{cc}
                                                \mathbf{A} & \mathbf{C} \\
                                                \mathbf{C}^H & \mathbf{B}
                                              \end{array}
\right |=|\mathbf{A}|\cdot |\mathbf{B}-\mathbf{C}^H\mathbf{A}^{-1}\mathbf{C}|$. Also recall the facts $|\mathbf{I}+\mathbf{M}_1\mathbf{M}_2|=|\mathbf{I}+\mathbf{M}_2\mathbf{M}_1|$  and $\mathbf{M}_1(\mathbf{M}_2\mathbf{M}_1+\mathbf{I})^{-1}\mathbf{M}_3=
(\mathbf{M}_1\mathbf{M}_2+\mathbf{I})^{-1}\mathbf{M}_1\mathbf{M}_3$ for compatible matrices.

Then
\begin{align}\label{eq:IYAYB}
&I(\mathbf{Y}_A^{(1)};\mathbf{Y}_B^{(1)})
=-\log |\mathbf{I}_{n_A\phi_A}-\mathbf{B}^{-1}\mathbf{C}^H\mathbf{A}^{-1}\mathbf{C}|.
\end{align}
Here
\begin{align}
&\mathbf{C}^H\mathbf{A}^{-1}\mathbf{C}=|\rho|^2\gamma_{BA}\gamma_{AB}
(\mathbf{\Pi}_B^*\otimes\mathbf{\Pi}_A^T)
\notag\\
&\,\,\cdot(\gamma_{AB}(\mathbf{\Pi}_B^T\mathbf{\Pi}_B^*\otimes\mathbf{I}_{n_A})+\mathbf{I}_{n_A\phi_B})^{-1}
(\mathbf{\Pi}_B^T\otimes\mathbf{\Pi}_A^*)\notag\\
&=|\rho|^2\gamma_{BA}\gamma_{AB}
(\mathbf{I}_{n_B}\otimes\mathbf{\Pi}_A^T)
\notag\\
&\,\,\cdot(\gamma_{AB}(\mathbf{\Pi}_B^*\mathbf{\Pi}_B^T\otimes\mathbf{I}_{n_A})
+\mathbf{I}_{n_An_B})^{-1}(\mathbf{\Pi}_B^*\mathbf{\Pi}_B^T\otimes\mathbf{\Pi}_A^*)\notag\\
&=|\rho|^2\frac{\gamma_{BA}\gamma_{AB}\psi_B}{\gamma_{AB}\psi_B+1}
(\mathbf{I}_{n_B}\otimes\mathbf{\Pi}_A^T\mathbf{\Pi}_A^*).
\end{align}
Then, from \eqref{eq:IYAYB},
\begin{align}\label{eq:IYAYB2}
&I(\mathbf{Y}_A^{(1)};\mathbf{Y}_B^{(1)})\notag\\
&=-\log\left |\mathbf{I}_{n_A\phi_B}-|\rho|^2(\gamma_{BA}(\mathbf{I}_{n_B}
\otimes\mathbf{\Pi}_A^T\mathbf{\Pi}_A^*)+\mathbf{I}_{n_B\phi_A})^{-1}\right .\notag\\
&\,\,\left . \cdot\frac{\gamma_{BA}\gamma_{AB}\psi_B}{\gamma_{AB}\psi_B+1}
(\mathbf{I}_{n_B}\otimes\mathbf{\Pi}_A^T\mathbf{\Pi}_A^*)\right |\notag\\
&=-\log\left |\mathbf{I}_{n_An_B}-|\rho|^2(\gamma_{BA}(\mathbf{I}_{n_B}
\otimes\mathbf{\Pi}_A^*\mathbf{\Pi}_A^T)+\mathbf{I}_{n_Bn_A})^{-1}\right .\notag\\
&\,\,\left . \cdot\frac{\gamma_{BA}\gamma_{AB}\psi_B}{\gamma_{AB}\psi_B+1}
(\mathbf{I}_{n_B}\otimes\mathbf{\Pi}_A^*\mathbf{\Pi}_A^T)\right |\notag
\\
&=-\log\left |\mathbf{I}_{n_An_B}-|\rho|^2\frac{\gamma_{BA}\gamma_{AB}\psi_B\psi_A}{(\gamma_{AB}\psi_B+1)
(\gamma_{BA}\psi_A+1)}
\mathbf{I}_{n_An_B}\right |\notag\\
&=n_An_B\log g.
\end{align}

\subsection{Proof of Theorem \ref{Theorem_a}}\label{proof_of_theorem_a}

\subsubsection{Analysis of $h(\mathcal{Y}|\mathcal{X})$}\label{hYX}
We can write by applying chain rule:
\begin{align}\label{eq:hYX_prim}
    \nonumber h(\mathcal{Y}|\mathcal{X})&=h(\mathbf{X}_B,\mathbf{Y}_{B}^{(1)},\mathbf{Y}_{B}^{(2)}|\mathbf{X}_A,\mathbf{Y}_{A}^{(1)},\mathbf{Y}_{A}^{(2)})\\
    \nonumber&=h(\mathbf{X}_B|\mathbf{X}_A,\mathbf{Y}_{A}^{(1)},\mathbf{Y}_{A}^{(2)})\\
    &\;+h(\mathbf{Y}_{B}^{(1)},\mathbf{Y}_{B}^{(2)}|\mathbf{X}_B,\mathbf{X}_A,\mathbf{Y}_{A}^{(1)},\mathbf{Y}_{A}^{(2)})
\end{align}
%
%
%
%
Here $\mathbf{X}_A$ is independent of $\{\mathbf{X}_B,\mathbf{Y}_{A}^{(1)},\mathbf{Y}_{A}^{(2)}\}$. And for large $\psi_B$, the condition on given $\mathbf{Y}_A^{(1)}$ is the same as the condition on given $\mathbf{H}_{AB}$ because of Lemma \ref{Lemma_high_power}. Hence, the 1st term in \eqref{eq:hYX_prim} is
\begin{align}\label{eq:hYX_a1}
&T_{\eqref{eq:hYX_prim},1}
    \approx h(\mathbf{X}_B|\mathbf{H}_{AB},\mathbf{Y}_{A}^{(2)})\notag\\
    &=h(\mathbf{X}_B)+h(\mathbf{Y}_{A}^{(2)}|\mathbf{H}_{AB},
    \mathbf{X}_B)-h(\mathbf{Y}_{A}^{(2)}|\mathbf{H}_{AB})
\end{align}
where the second equation uses the fact $h(A,B|C)=h(A|C)+h(B|A,C)=h(B|C)+h(A|B,C)$. Also note that $h(\mathbf{X}_B|\mathbf{H}_{AB})=h(\mathbf{X}_B)$. Such a technique will be used frequently without further explanation.

We can write the 2nd term in \eqref{eq:hYX_prim} as:
\begin{align}\label{eq:hYX_b1}
     &T_{\eqref{eq:hYX_prim},2}= h(\mathbf{Y}_{B}^{(1)}|\mathbf{X}_B, \mathbf{X}_A,  \mathbf{Y}_{A}^{(1)} , \mathbf{Y}_{A}^{(2)})\notag\\
    &\qquad+h(\mathbf{Y}_{B}^{(2)}|\mathbf{Y}_{B}^{(1)} ,\mathbf{X}_B,\mathbf{X}_A, \mathbf{Y}_{A}^{(1)} , \mathbf{Y}_{A}^{(2)}).
\end{align}

For the first term in \eqref{eq:hYX_b1}, $\mathbf{X}_A$ is independent of $\{\mathbf{Y}_B^{(1)},\mathbf{X}_B,\mathbf{Y}_{A}^{(1)},\mathbf{Y}_{A}^{(2)}\}$. So we can write,
\begin{align}
      &T_{\eqref{eq:hYX_b1},1}=h(\mathbf{Y}_{B}^{(1)}|\mathbf{X}_B, \mathbf{Y}_{A}^{(1)},\mathbf{Y}_{A}^{(2)})\notag\\
      &\approx h(\mathbf{Y}_{B}^{(1)}|\mathbf{X}_B, \mathbf{Y}_{A}^{(1)},\mathbf{H}_{A,B},\mathbf{Y}_{A}^{(2)})\notag\\
      &=h(\mathbf{Y}_{B}^{(1)}|\mathbf{Y}_{A}^{(1)},\mathbf{H}_{A,B})\approx h(\mathbf{Y}_{B}^{(1)}|\mathbf{Y}_{A}^{(1)}).
\end{align}
where the approximations are due to large $\psi_B$, and the 3rd line is due to independence between $\{\mathbf{Y}_{B}^{(1)},\mathbf{Y}_{A}^{(1)}\}$ and $\{\mathbf{X}_B,\mathbf{Y}_{A}^{(2)}\}$ when $\mathbf{H}_{A,B}$ is given.

For the second term in \eqref{eq:hYX_b1}, we have
\begin{align}\label{eq:hYX_b1.1}
     & T_{\eqref{eq:hYX_b1},2}\approx h(\mathbf{Y}_{B}^{(2)}|\mathbf{H}_{BA} ,\mathbf{X}_B,\mathbf{X}_A, \mathbf{H}_{AB}, \mathbf{Y}_{A}^{(2)})\notag\\
     &= h(\mathbf{Y}_{B}^{(2)}|\mathbf{H}_{BA},\mathbf{X}_A).
\end{align}
where the approximation is due to large $\psi_A$ and $\psi_B$, and the second equation is because given $\{\mathbf{H}_{BA},\mathbf{X}_A\}$, $\mathbf{Y}_{B}^{(2)}$ is independent of $\{\mathbf{X}_B,\mathbf{H}_{AB},\mathbf{Y}_A^{(2)}\}$.

Using the above  results for large $\psi_A$ and $\psi_B$, \eqref{eq:hYX_prim} becomes
\begin{align}\label{eq:hYX_fin}
    \nonumber h(\mathcal{Y}|\mathcal{X})&\approx h(\mathbf{X}_B)+h(\mathbf{Y}_{A}^{(2)}|\mathbf{H}_{AB},\mathbf{X}_B)-
    h(\mathbf{Y}_{A}^{(2)}|\mathbf{H}_{AB})\\
    &+h(\mathbf{Y}_{B}^{(1)}|\mathbf{Y}_{A}^{(1)})+h(\mathbf{Y}_{B}^{(2)}|\mathbf{H}_{BA},\mathbf{X}_A).
\end{align}
Note that the above decomposition of $h(\mathcal{Y}|\mathcal{X})$ is such that the closed-form expression of each component can be found directly from the data model. The same objective is applied to $h(\mathcal{Y}|\mathcal{Z})$, $h(\mathcal{X}|\mathcal{Z})$ and $h(\mathcal{X}|\mathcal{Y},\mathcal{Z})$ next.
\subsubsection{Analysis of $h(\mathcal{Y}|\mathcal{Z})$ and $h(\mathcal{X}|\mathcal{Z})$}\label{hYZ}
We can write
\begin{align}\label{eq:hYZ_prim}
    \nonumber &h(\mathcal{Y}|\mathcal{Z})=h(\mathbf{X}_B,\mathbf{Y}_{B}^{(1)},\mathbf{Y}_{B}^{(2)}|
    \mathbf{Y}_{EA}^{(1)},\mathbf{Y}_{EA}^{(2)},\mathbf{Y}_{EB}^{(1)},\mathbf{Y}_{EB}^{(2)})\\
    \nonumber&=h(\mathbf{X}_B|\mathbf{Y}_{EA}^{(1)},\mathbf{Y}_{EA}^{(2)},\mathbf{Y}_{EB}^{(1)},
    \mathbf{Y}_{EB}^{(2)})\\
    &\,\,+h(\mathbf{Y}_{B}^{(1)},\mathbf{Y}_{B}^{(2)}|\mathbf{X}_B,\mathbf{Y}_{EA}^{(1)},
    \mathbf{Y}_{EA}^{(2)},\mathbf{Y}_{EB}^{(1)},\mathbf{Y}_{EB}^{(2)}).
\end{align}
Here we see that the first term in \eqref{eq:hYZ_prim} is
\begin{align}
&T_{\eqref{eq:hYZ_prim},1}\approx h(\mathbf{X}_B|\mathbf{G}_A,\mathbf{Y}_{EA}^{(2)},\mathbf{G}_B,
    \mathbf{Y}_{EB}^{(2)})\notag\\
    &=h(\mathbf{X}_B|\mathbf{G}_B,
    \mathbf{Y}_{EB}^{(2)})\notag\\
    &=h(\mathbf{X}_B)+h(\mathbf{Y}_{EB}^{(2)}|\mathbf{X}_B,\mathbf{G}_B)
    -h(\mathbf{Y}_{EB}^{(2)}|\mathbf{G}_B)
\end{align}
where the approximation is due to large $\psi_A$ and $\psi_B$, and the second equation is because of independence between the conditioning matrices $\{\mathbf{G}_A,\mathbf{Y}_{EA}^{(2)}\}$ and the other matrices $\{\mathbf{X}_B,\mathbf{G}_B,
    \mathbf{Y}_{EB}^{(2)}\}$.
%

Furthermore, the second term in \eqref{eq:hYZ_prim} is
\begin{align}
&T_{\eqref{eq:hYZ_prim},2}=h(\mathbf{Y}_{B}^{(1)},\mathbf{Y}_{B}^{(2)}|\mathbf{Y}_{EA}^{(1)},
    \mathbf{Y}_{EA}^{(2)})\notag\\
    &=h(\mathbf{Y}_{B}^{(1)}|\mathbf{Y}_{EA}^{(1)},
    \mathbf{Y}_{EA}^{(2)})+h(\mathbf{Y}_{B}^{(2)}|\mathbf{Y}_{B}^{(1)},\mathbf{Y}_{EA}^{(1)},
    \mathbf{Y}_{EA}^{(2)})\notag\\
    &=h(\mathbf{Y}_{B}^{(1)})+h(\mathbf{Y}_{B}^{(2)},\mathbf{Y}_{EA}^{(2)}|
    \mathbf{Y}_{B}^{(1)},\mathbf{Y}_{EA}^{(1)})\notag\\
    &\,\,-
    h(\mathbf{Y}_{EA}^{(2)}|\mathbf{Y}_{B}^{(1)},\mathbf{Y}_{EA}^{(1)})\notag\\
    &\approx h(\mathbf{Y}_{B}^{(1)})+h(\mathbf{Y}_{B}^{(2)},\mathbf{Y}_{EA}^{(2)}|
    \mathbf{H}_{BA},\mathbf{G}_{A})\notag\\
    &\,\,-
    h(\mathbf{Y}_{EA}^{(2)}|\mathbf{G}_{A})
\end{align}
where the first equation is due to independence between the conditioning matrices $\{\mathbf{X}_B,
  \mathbf{Y}_{EB}^{(1)},\mathbf{Y}_{EB}^{(2)}\}$ and the other matrices $\{\mathbf{Y}_{B}^{(1)},\mathbf{Y}_{B}^{(2)},\mathbf{Y}_{EA}^{(1)},\mathbf{Y}_{EA}^{(2)}\}$, the second and third equations applied the chain rule, and the last approximation is due to larger $\psi_A$ and $\psi_B$. All dropped conditioning matrices are due to independence.

Combining the above results for large $\psi_A$ and $\psi_B$,  \eqref{eq:hYZ_prim} becomes
\begin{align}\label{eq:hYZ_fin}
    \nonumber &h(\mathcal{Y}|\mathcal{Z}) \approx h(\mathbf{X}_B ) + h(\mathbf{Y}_{EB}^{(2)}|\mathbf{G}_B, \mathbf{X}_B ) - h(\mathbf{Y}_{EB}^{(2)}|\mathbf{G}_B ) \\
    &\,\,+ h(\mathbf{Y}_{B}^{(1)})+ h(\mathbf{Y}_{EA}^{(2)},\mathbf{Y}_{B}^{(2)}|\mathbf{G}_A,\mathbf{H}_{BA}) \notag\\
    &\,\,- h(\mathbf{Y}_{EA}^{(2)}|\mathbf{G}_A).
\end{align}
By symmetry between $h(\mathcal{Y}|\mathcal{Z})$ and $h(\mathcal{X}|\mathcal{Z})$, it follows from \eqref{eq:hYZ_fin} that for large $\psi_A$ and $\psi_B$,
\begin{align}\label{eq:hXZ_fin}
    \nonumber &h(\mathcal{X}|\mathcal{Z}) \approx h(\mathbf{X}_A) + h(\mathbf{Y}_{ E A}^{(2)}|\mathbf{G}_A, \mathbf{X}_A) - h(\mathbf{Y}_{ E A}^{(2)}|\mathbf{G}_A) \\
    &\,\,+ h(\mathbf{Y}_{A}^{(1)})+ h(\mathbf{Y}_{EB}^{(2)},\mathbf{Y}_{A}^{(2)}|\mathbf{G}_B,\mathbf{H}_{AB}) \notag\\
    &\,\,- h(\mathbf{Y}_{EB}^{(2)}|\mathbf{G}_B).
\end{align}
\subsubsection{Analysis of $h(\mathcal{X}|\mathcal{Y},\mathcal{Z})$ and $h(\mathcal{Y}|\mathcal{X},\mathcal{Z})$}\label{hXYZ}
We have
\begin{align}\label{eq:hXYZ_prim}
    \nonumber&h(\mathcal{X}|\mathcal{Y},\mathcal{Z})=h(\mathbf{X}_A, \mathbf{Y}_A^{(1)}, \mathbf{Y}_A^{(2)}|\mathcal{Y},\mathcal{Z})\\
    \nonumber&=h(\mathbf{X}_A|\mathcal{Y},\mathcal{Z})+h(\mathbf{Y}_A^{(1)}, \mathbf{Y}_A^{(2)}|\mathbf{X}_A,\mathcal{Y},\mathcal{Z})\\
    \nonumber&=h(\mathbf{X}_A|\mathbf{Y}_B^{(1)},\mathbf{Y}_B^{(2)},\mathbf{Y}_{EA}^{(1)},\mathbf{Y}_{EA}^{(2)})\\
    &\,\,+h(\mathbf{Y}_A^{(1)},\mathbf{Y}_A^{(2)}|\mathbf{X}_A,
    \mathbf{X}_B,\mathbf{Y}_B^{(1)},\mathbf{Y}_B^{(2)}).
\end{align}
Here we have used the fact that $\{\mathbf{X}_B,\mathbf{Y}_{EB}^{(1)},\mathbf{Y}_{EB}^{(2)}\}$ is independent of $\{\mathbf{X}_A,\mathbf{Y}_B^{(1)},\mathbf{Y}_B^{(2)},\mathbf{Y}_{EA}^{(1)},
\mathbf{Y}_{EA}^{(2)}\}$; and given $\{\mathbf{X}_A,\mathbf{X}_B\}$, $\{\mathbf{Y}_{EA}^{(1)},\mathbf{Y}_{EA}^{(2)},\mathbf{Y}_{EB}^{(1)},\mathbf{Y}_{EB}^{(2)}\}$ is independent of $\{\mathbf{Y}_A^{(1)},\mathbf{Y}_A^{(2)},\mathbf{Y}_B^{(1)},\mathbf{Y}_B^{(2)}\}$.

The first term in  \eqref{eq:hXYZ_prim} for large $\psi_A$ is
\begin{align}\label{eq:hXYZ_a1}
    \nonumber&
    T_{\eqref{eq:hXYZ_prim},1}\approx h(\mathbf{X}_A|\mathbf{H}_{BA},\mathbf{G}_A, \mathbf{Y}_B^{(2)} , \mathbf{Y}_{EA}^{(2)})\\
    \nonumber&=h(\mathbf{X}_A)+h(\mathbf{Y}_{EA}^{(2)},\mathbf{Y}_{B}^{(2)}|\mathbf{G}_A,\mathbf{H}_{BA},\mathbf{X}_A)\\
    &\,\,-h(\mathbf{Y}_{EA}^{(2)},\mathbf{Y}_{B}^{(2)}|\mathbf{G}_A,\mathbf{H}_{BA}).
\end{align}
The second term in \eqref{eq:hXYZ_prim}  is
\begin{align}
&T_{\eqref{eq:hXYZ_prim},2}=h(\mathbf{Y}_A^{(1)}|\mathbf{X}_A,
    \mathbf{X}_B,\mathbf{Y}_B^{(1)},\mathbf{Y}_B^{(2)})\notag\\
    &\,\,
    +h(\mathbf{Y}_A^{(2)}|\mathbf{Y}_A^{(1)},\mathbf{X}_A,
    \mathbf{X}_B,\mathbf{Y}_B^{(1)},\mathbf{Y}_B^{(2)})\notag\\
    &\approx h(\mathbf{Y}_A^{(1)}|\mathbf{Y}_B^{(1)})+h(\mathbf{Y}_A^{(2)}|\mathbf{H}_{AB},\mathbf{X}_A,
    \mathbf{X}_B,\mathbf{H}_{BA},\mathbf{Y}_B^{(2)})\notag\\
    &=h(\mathbf{Y}_A^{(1)}|\mathbf{Y}_B^{(1)})+h(\mathbf{Y}_A^{(2)}|\mathbf{H}_{AB},
    \mathbf{X}_B)
\end{align}
where the approximation is due to large $\psi_A$ and $\psi_B$.

Therefore, for large $\psi_A$ and $\psi_B$, \eqref{eq:hXYZ_prim} becomes
\begin{align}\label{eq:hXYZ_fin}
    \nonumber&h(\mathcal{X}|\mathcal{Y},\mathcal{Z}) \approx h(\mathbf{Y}_{EA}^{(2)}, \mathbf{Y}_{B}^{(2)}|\mathbf{G}_{A},\mathbf{H}_{BA},\mathbf{X}_A) + h(\mathbf{Y}_{A}^{(1)}|\mathbf{Y}_B^{(1)})\\
    &\,\,+ h(\mathbf{X}_A ) - h(\mathbf{Y}_{ EA}^{(2)}, \mathbf{Y}_{B}^{(2)}|\mathbf{G}_A, \mathbf{H}_{ BA} ) \notag\\
    &\,\,+ h(\mathbf{Y}_A^{(2)}|\mathbf{H}_{ AB},\mathbf{X}_B ).
\end{align}
%
%
By symmetry, it follows from \eqref{eq:hXYZ_fin} that for large $\psi_A$ and $\psi_B$,
\begin{align}\label{eq:hYXZ_fin}
    \nonumber&h(\mathcal{Y}|\mathcal{X},\mathcal{Z}) \approx h(\mathbf{Y}_{EB}^{(2)}, \mathbf{Y}_{A}^{(2)}|\mathbf{G}_{B},\mathbf{H}_{AB},\mathbf{X}_B) + h(\mathbf{Y}_{B}^{(1)}|\mathbf{Y}_A^{(1)})\\
    &\,\,+ h(\mathbf{X}_B ) - h(\mathbf{Y}_{ EB}^{(2)}, \mathbf{Y}_{A}^{(2)}|\mathbf{G}_B, \mathbf{H}_{ AB} ) \notag\\
    &\,\,+ h(\mathbf{Y}_B^{(2)}|\mathbf{H}_{ BA},\mathbf{X}_A ).
\end{align}
\subsubsection{Proof of Theorem \ref{Theorem_a}}
It follows from \eqref{eq:hYX_fin} and \eqref{eq:hYZ_fin} that for large $\psi_A$ and $\psi_B$,
\begin{align}\label{eq:C_B_a}
    &C_B=h(\mathcal{Y}|\mathcal{Z})-h(\mathcal{Y}|\mathcal{X})\notag\\
    &\approx h(\mathbf{Y}_{B}^{(1)})-h(\mathbf{Y}_{B}^{(1)}|\mathbf{Y}_A^{(1)})\notag\\
    &\,\,+h(\mathbf{Y}_{EB}^{(2)}|\mathbf{G}_{B},\mathbf{X}_B)-
    h(\mathbf{Y}_{EB}^{(2)}|\mathbf{G}_{B})\notag\\
    &\,\,+h(\mathbf{Y}_{A}^{(2)}|\mathbf{H}_{AB})
    -h(\mathbf{Y}_{A}^{(2)}|\mathbf{H}_{AB},\mathbf{X}_B)\notag\\
    &\,\,+h(\mathbf{Y}_{EA}^{(2)},\mathbf{Y}_{B}^{(2)}|\mathbf{G}_A,\mathbf{H}_{BA})
    -h(\mathbf{Y}_{EA}^{(2)}|\mathbf{G}_A)\notag\\
    &\,\,-h(\mathbf{Y}_{B}^{(2)}|\mathbf{H}_{BA},\mathbf{X}_A).
\end{align}
We see that the first two terms in \eqref{eq:C_B_a} is
\begin{equation}\label{}
  T_{\eqref{eq:C_B_a},1,2}=I(\mathbf{Y}_{A}^{(1)};\mathbf{Y}_{B}^{(1)})=C_S^{(1)}
\end{equation}
which is given in Lemma \ref{Lemma_MI}. Similarly,
\begin{align}\label{}
  &T_{\eqref{eq:C_B_a},3,4}=-I(\mathbf{Y}_{EB}^{(2)};\mathbf{X}_B|\mathbf{G}_{B})\notag\\
  &=-v_B\mathbb{E}\{\log|\mathbf{I}_{n_E}+\gamma_{EB}\mathbf{G}_B\mathbf{G}_B^H|\},\\
&T_{\eqref{eq:C_B_a},5,6}=I(\mathbf{Y}_A^{(2)};\mathbf{X}_B|\mathbf{H}_{AB})\notag\\
&=v_B\mathbb{E}\{\log|\mathbf{I}_{n_A}+\gamma_{AB}\mathbf{H}_{AB}\mathbf{H}_{AB}^H|\}.
\end{align}

For the 7th term in \eqref{eq:C_B_a}, we now rewrite
 \eqref{y_B} and \eqref{y_EA} as follows,
\begin{align}\label{eq:H1}
    \begin{bmatrix}\mathbf{Y}_{B}^{(2)}\\\mathbf{Y}_{EA}^{(2)}\end{bmatrix}&
    =\begin{bmatrix}\sqrt{\gamma_{BA}}\mathbf{H}_{BA}\\\sqrt{\gamma_{EA}}
    \mathbf{G}_{A}\end{bmatrix}\mathbf{X}_A+\begin{bmatrix}\mathbf{W}_{B}^{(2)}\\
    \mathbf{W}_{EA}^{(2)}\end{bmatrix}
\end{align}
for which we will also write
\begin{align}\label{}
  &\sqrt{\gamma_{EA}}\tilde {\mathbf{G}}_A\doteq \left[ \begin{array}{c}
                                                   \sqrt{\gamma_{BA}}\mathbf{H}_{BA} \\
                                                   \sqrt{\gamma_{EA}}
    \mathbf{G}_{A}
                                                 \end{array}
  \right ]=\sqrt{\gamma_{EA}}
  \left [\begin{array}{c}
      \sqrt{\frac{\lambda_{EA}}{\lambda_B}}\mathbf{H}_{BA} \\
                \mathbf{G}_{A}
                        \end{array}
  \right ].\notag
\end{align}

Then   applying Lemma \ref{Lemma_MI} to \eqref{eq:H1}, we obtain
\begin{align}
    \nonumber&T_{\eqref{eq:C_B_a},7}=(n_B+n_E)v_A\log(\pi e)\\
    &\,\,+v_A\mathbb{E}\{\log|\mathbf{I}_{n_B+n_E}+
    \gamma_{EA}\tilde{\mathbf{G}}_{A}\tilde{\mathbf{G}}_{A}^H|\},\\
  &T_{\eqref{eq:C_B_a},8}=-n_Ev_A\log(\pi e)\notag\\
  &\,\,-v_A\mathbb{E}\{\log|\mathbf{I}_{n_A}+\gamma_{EA}\mathbf{G}_A^H\mathbf{G}_A|\},\\
  &T_{\eqref{eq:C_B_a},9}=-n_Bv_A\log(\pi e).
\end{align}

Combining the above results, \eqref{eq:C_B_a} becomes
\begin{align}\label{eq:C_B_general}
&C_B=C_S^{(1)}-v_B\mathbb{E}\{\log|\gamma_{EB}\mathbf{G}_B\mathbf{G}_B^H+\mathbf{I}_{n_E}|\}\notag\\
&\,\,+v_B\mathbb{E}\{\log|\gamma_{AB}\mathbf{H}_{AB}\mathbf{H}_{AB}^H+\mathbf{I}_{n_A}|\}\notag\\
&\,\,+v_A\mathbb{E}\{\log|\gamma_{EA}\tilde{\mathbf{G}}_{A}\tilde{\mathbf{G}}_{A}^H
+\mathbf{I}_{n_B+n_E}|\}\notag\\
&\,\,-v_A\mathbb{E}\{\log|\gamma_{EA}\mathbf{G}_A\mathbf{G}_A^H+\mathbf{I}_{n_E}|\}.
\end{align}

Similar to the analysis of $C_B$, it follows  from \eqref{eq:hYX_fin} and \eqref{eq:hYXZ_fin} that for large $\psi_A$ and $\psi_B$,
\begin{align}\label{eq:diff}
&C_Z-C_B = h(\mathcal{Y}|\mathcal{X})-h(\mathcal{Y}|\mathcal{X},\mathcal{Z})\notag\\
&\approx
h(\mathbf{Y}_{A}^{(2)}|\mathbf{H}_{AB},\mathbf{X}_B)-
    h(\mathbf{Y}_{A}^{(2)}|\mathbf{H}_{AB})
    \notag\\
    &\,\,+h(\mathbf{Y}_{ EB}^{(2)}, \mathbf{Y}_{A}^{(2)}|\mathbf{G}_B, \mathbf{H}_{ AB})\notag\\
    &\,\,-
    h(\mathbf{Y}_{EB}^{(2)}, \mathbf{Y}_{A}^{(2)}|\mathbf{G}_{B},\mathbf{H}_{AB},\mathbf{X}_B)\notag\\
    &=-I(\mathbf{Y}_{A}^{(2)};\mathbf{X}_B|\mathbf{H}_{AB})\notag\\
    &\,\,+I(\{\mathbf{Y}_{ EB}^{(2)}, \mathbf{Y}_{A}^{(2)}\};\mathbf{X}_B|\mathbf{G}_{B},\mathbf{H}_{AB})\notag\\
    &=-v_B\mathbb{E}\{\log|\mathbf{I}_{n_A}+\gamma_{AB}\mathbf{H}_{AB}\mathbf{H}_{AB}^H|\}\notag\\
    &\,\,+v_B\mathbb{E}\{\log|\mathbf{I}_{n_A+n_E}+
    \gamma_{AB}\tilde{\mathbf{H}}_{AB}\tilde{\mathbf{H}}_{AB}^H|\}.
\end{align}
Here $\tilde{\mathbf{H}}_{AB}=[\mathbf{H}_{AB}^T,\sqrt{\lambda_A/\lambda_{EB}}\mathbf{G}_B^T]^T$.

A simple application of the above results completes the proof of Theorem \ref{Theorem_a}.

\end{appendices}



\end{document}